\documentclass[prl,aps,twocolumn,preprintnumbers,showpacs,showkeys,superscriptaddress]{revtex4}
\def\bfl{\begin{flushleft}}
\def\efl{\end{flushleft}}
\def\bfr{\begin{flushright}}
\def\efr{\end{flushright}}
\def\bc{\begin{center}}
\def\ec{\end{center}}
\def\be{\begin{equation}}
\def\ee{\end{equation}}
\def\ba{\begin{eqnarray}}
\def\ea{\end{eqnarray}}
\def\baa#1{\begin{array}{#1}}
\def\eaa{\end{array}}
\def\bw{\begin{widetext}}
\def\ew{\end{widetext}}

\def\lb#1{\label{#1}}

\def\text#1{\mbox{#1}}

\begin{document}


\title{Resistivity model in both paramagnetic and
ferromagnetic phases}

\author{Andrew Das Arulsamy}

\address{Condensed Matter Group, Division of Exotic Matter, No. 22, Jalan Melur 14,
Taman Melur, 68000 Ampang, Selangor DE, Malaysia}


\date{\today}

\keywords{Ferromagnets, Resistivity model, Ionization energy,
Fermi-Dirac statistics}


\begin{abstract}

The resistivity model as a function of temperature and ionization
energy (doping) is derived with further confinements from
spin-disorder scattering in ferromagnetic phase. Magnetization and
polaronic effects capture the mechanism of both spin independent
and spin-assisted charge transport of ferromagnets. The computed
$T_{crossover}$ below $T_C$ and carrier density in
Ga$_{1-x}$Mn$_x$As system are 8-12 K and 10$^{19}$ cm$^{-3}$,
remarkably identical with the experimental values of 10-12 K and
10$^{18}$-10$^{20}$ cm$^{-3}$ respectively.
\end{abstract}

\pacs{72.10.Bg; 71.30.+h; 71.38.Ht; 75.50.Pp}

\maketitle



\section{1. Introduction}\lb{s-in}

Diluted magnetic semiconductors (DMS) have the tremendous
potential for the development of spintronics and subsequently will
lay the foundation to realize quantum computing. This
applicability arises due to ferromagnetic nature of DMS. In other
words, both the charge and spin of the electrons can be exploited
with limited Mn doping in GaAs semiconductor. In order to achieve
this, one needs to understand the transport mechanism such as the
variation of resistivity with temperature and doping in both above
and below $T_C$ consistently. Interestingly, Van Esch {\it et
al}.~\cite{esch1} have proposed multiple exchange interactions,
which are ferromagnetic (FM) hole-hole and antiferromagnetic (AFM)
Mn-hole interactions for DMS. These two effects, after neglecting
the direct exchange between Mn-Mn (due to very diluted nature of
DMS) are seem to be sufficient enough to describe the temperature
dependent magnetization curves ($M(T)$) accurately. However, even
after inclusion of FM and AFM effects including the spin disorder
scattering, the transport property in the FM phase is still not
well understood. Unfortunately, this is also true for the case of
metallic property below $T_C$ in the well known and extensively
studied FM manganites as pointed out by Mahendiran {\it et
al}.~\cite{mahendiran2}. The resistivity ($\rho(T)$) above $T_C$
for manganites is found to be in an activated form described by
the equation~\cite{mahendiran2},

\begin {eqnarray}
\rho(T>T_C) = \rho_0\exp\bigg(\frac{E_a}{k_BT}\bigg).\label{eq:1}
\end {eqnarray}

$E_a$ is the activation energy, $\rho_0$ and $k_B$ denote the
residual resistivity at $T$ $\gg$ $E_a$ and Boltzmann constant
respectively. In the FM phase, the influence of $M(T)/M_{4.2}$ is
more pronounced than the electron-phonon ({\it e-ph}) contribution
where the latter requires an overwhelmingly large coupling
constant~\cite{mahendiran2}. Note that $M_{4.2}$ is the magnitude
of magnetization at 4.2 K. Therefore, Mahendiran {\it et al}. have
suggested that conventional mechanism namely, {\it e-ph}
scattering has to be put aside so as to explain the $\rho(T)$ for
manganites below $T_C$. On the contrary, $\rho(T)$ with {\it e-ph}
involvement for DMS in the paramagnetic phase is given
by~\cite{esch1}

\begin {eqnarray}
\rho(T>T_C) = \frac{C_1 +
C_2\big[\exp\big(\Theta_D/T\big)-1\big]^{-1}}{k_BT\ln\big[1 +
\exp\big((E_m - E_f)/k_BT\big)\big]}.\label{eq:2}
\end {eqnarray}

The term, $C_2/\big[\exp(\Theta_D/T)-1\big]$ takes care of the
{\it e-ph} contribution. $\Theta_D$, $E_f$, $E_m$, $C_1$ and $C_2$
represent the Debye temperature, Fermi level, mobility edge and
numerical constants respectively. The $\rho(T)$ in the FM phase
based on the spin disorder scattering as derived by
Tinbergen-Dekker is given by~\cite{tinbergen3}

\begin {eqnarray}
&&\rho_{SD}(T<T_C) = \frac{(m^*_{e,h})^{5/2}N(2E_F)^{1/2}}{\pi
(n,p) e^2\hbar^4}J_{ex}^2 \nonumber \\&& \times \bigg[S(S+1) -
S^2\bigg(\frac{M_{TD}(T)}{M_{4.2}}\bigg)^2  \nonumber \\&& -
S\bigg(\frac{M_{TD}(T)}{M_{4.2}}\bigg)\tanh
\bigg(\frac{3T_CM_{TD}(T)}{2TS(S+1)M_{4.2}}\bigg)\bigg].\label{eq:3}
\end {eqnarray}

$N$ is the concentration of nearest neighbor ions (Mn's
concentration) while ($n,p$) is the concentration of charge
carriers (electrons or holes respectively). $m_{e,h}^*$ denotes
effective mass of electrons or holes, $\hbar$ = $h/2\pi$, $h$ =
Planck constant. $e$ is the charge of an electron, $E_F$ and
$J_{ex}$ are the Fermi and FM exchange interaction energies
respectively while $S$ is the spin quantum number.
Equation~(\ref{eq:3}) becomes equivalent to Kasuya~\cite{kasuya4}
if one replaces the term,
$\tanh\big[3T_CM_{TD}(T)/2TS(S+1)M_{4.2}\big]$ with 1. Again, an
accurate equation for the $\rho(T)$ below $T_C$ is still lacking
since spin disorder scattering alone is insufficient as shown by
Tinbergen and Dekker~\cite{tinbergen3} as well as reviewed by
Ohno~\cite{ohno5}. Hence, it is desirable to derive a formula that
could describe the transport mechanism of ferromagnets for the
whole temperature range i.e., for both paramagnetic and FM phases
and even at very low $T$. With this in mind, the ionization energy
based Fermi-Dirac statistics (iFDS) as derived in
Ref.~\cite{arulsamy6,arulsamy7,arulsamy8,arulsamy9,arulsamy10} and
spin disorder scattering based resistivity models derived by both
Tinbergen-Dekker and Kasuya will be employed in order to derive
$\rho$ as a function of $T$, ionization energy ($E_I$) and
$M_{\rho}(T,M_{4.2})$. The consequences of
$\rho(T,E_I,M_{\rho}(T,M_{4.2}))$ that arises from the variation
of $T$, $E_I$ and $M_{\rho}(T,M_{4.2})$ are discussed in detail
based on the experimental data reported by Van Esch {\it et
al}.~\cite{esch1} and Mahendiran {\it et al}.~\cite{mahendiran2}.

\section{2. Resistivity model}\lb{s-in}

The total current in semiconducting ferromagnets with
contributions from both paramagnetic and FM phases is $J$ =
$\sum_\nu J_\nu$, $\nu$ = $e^{\downarrow}$, $se^{\uparrow}$,
$h^{\downarrow}$, $sh^{\uparrow}$. For convenience, the spin-up,
$\uparrow$ denotes the direction of the magnetic field or a
particular direction below $T_C$, while the spin-down,
$\downarrow$ represents any other directions. Note that the total
energy (Kinetic + Magnetic), $E_{K+M}^{\uparrow}$ $\neq$
$E_{K+M}^{\downarrow}$ due to energy level splitting below $T_C$.
As such, the total current can be simplified as $J$ =
$J^{\downarrow}_e$ + $J^{\uparrow}_{se}$ = $J_e$ + $J_{se}$ if the
considered system is a $n$-type while $J$ = $J_h$ + $J_{sh}$ if it
is a $p$-type. $J_e$ and $J_h$ are the spin independent charge
current (electrons and holes respectively) in the paramagnetic
phase whereas $J_{se}$ and $J_{sh}$ are the spin-assisted charge
current in the FM phase. Thus the total resistivity ($n$ or
$p$-type) can be written as

\begin {eqnarray}
\rho^{-1} &&= \rho^{-1}_{e,h} + \rho^{-1}_{se,sh} \nonumber
\\&& =\bigg[\frac{m_{e,h}^*}{(n,p)e^2\tau_e}\bigg]^{-1} +
\bigg[\frac{m_{e,h}^*}{(n,p)e^2\tau_{SD}}\bigg]^{-1}.\label{eq:4}
\end {eqnarray}

$\tau_{SD}$ represents the spin disorder scattering rate. The
carrier density for the electrons and holes ($n,p$) based on iFDS
are given
by~\cite{arulsamy6,arulsamy7,arulsamy8,arulsamy9,arulsamy10}

\begin{eqnarray}
n = 2\left[\frac{k_BT}{2\pi\hbar^2}\right]^{3/2}(m^*_e)^{3/2}
\exp\left[\frac{E_F - E_I}{k_BT}\right]. \label{eq:5}
\end{eqnarray}

\begin{eqnarray}
p = 2\left[\frac{k_BT}{2\pi\hbar^2}\right]^{3/2}(m^*_h)^{3/2}
\exp\left[\frac{-E_F - E_I}{k_BT}\right]. \label{eq:6}
\end{eqnarray}

The derivation of iFDS, $f(E_I) =
\exp\big[-\mu-\lambda(E_{initial~state} \pm E_I)\big]$ by
employing the restrictive conditions, $\sum_i^{\infty} dn_i = 0$
and $\sum_i^{\infty} (E_{initial~state}\pm E_I)_i dn_i = 0$ is
well documented (including its applications) in the
Refs.~\cite{arulsamy6,andrew,arulsamy7,arulsamy8,arulsamy9,arulsamy10,andrew2,andrew3}.
$E_{initial~state}$ denotes the energy at certain initial state
and the Lagrange multipliers, $\mu_e + \lambda E_I =
-\ln\big[(n/V)(2\pi\lambda\hbar^2/m_e)^{3/2}\big]$, $\mu_h -
\lambda E_I = \ln\big[(p/V)(2\pi\lambda\hbar^2/m_h)^{3/2}\big]$
and $\lambda$ = 1/$k_BT$. $V$ is the volume in {\bf k} space and
the gap-parameter, $E_I$ that represents electron-ion attraction
is also a parameter that measures the combination of electrons and
its strain field due to neighboring ions, which is nothing but
polarons~\cite{arulsamy6,arulsamy66}. The absolute value of $E_I$
can be obtained from~\cite{arulsamy6}, $E_I
=e^2/8\pi\epsilon\epsilon_0r_B$. $\epsilon$ and $\epsilon_0$ are
the dielectric constant and permittivity of free space
respectively, $r_B$ is the Bohr radius. Furthermore, the variation
of $E_I$ with magnetic field, {\bf H} will give rise to an inverse
variation on $r_B$ that also takes care of the polaronic
effect~\cite{arulsamy6,arulsamy66}. Substituting $1/\tau_e$ =
$AT^2$ (due to electron-electron interaction), Eqs.~(\ref{eq:3})
and~(\ref{eq:5}) or~(\ref{eq:6}) into Eq.~(\ref{eq:4}), then one
can arrive at

\begin{eqnarray}
\rho_{e,se}(T) = \frac{AB\exp
\big[(E_I+E_F)/k_BT\big]}{AT^{3/2}[M_{\rho}(T,M_{4.2})]^{-1}+
BT^{-1/2}}. \label{eq:7}
\end{eqnarray}

In which, $A =
[A_{e,h}/2e^2(m^*_{e,h})^{1/2}][2\pi\hbar^2/k_B]^{3/2}$, $B =
2m^*_{e,h}N(\pi E_F)^{1/2}J_{ex}^2/e^2\hbar k_B^{3/2}$ and
$\tau_{SD}^{-1} = [N(2E_F)^{1/2}(m^*_{e,h})^{3/2}/\pi
\hbar^4]J_{ex}^2M_{\rho}(T,M_{4.2})$. $A_{e,h}$ is the $T$
independent electron-electron scattering rate constant. The
empirical function of the normalized magnetization is given by

\begin{eqnarray}
M_{\rho}(T,M_{4.2}) = 1-\frac{M_{\rho}(T)}{M_{4.2}}\label{eq:8}.
\end{eqnarray}

Equation~(\ref{eq:8}) is an empirical function that directly
quantifies the influence of spin alignments in the FM phase on the
transport properties of charge and spin carriers~\cite{arulsamy66}
in accordance with Eq.~(\ref{eq:7}). In other words, the only way
to obtain $\frac{M_{\rho}(T)}{M_{4.2}}$ is through
Eq.~(\ref{eq:8}). In fact, Eq.~(\ref{eq:8}) is used to calculate
$M_{TD}(T)/M_{4.2}$ and $M_{K}(T)/M_{4.2}$ by writing $S(S+1) -
S^2\big(\frac{M_{TD}(T)}{M_{4.2}}\big)^2 -
S\big(\frac{M_{TD}(T)}{M_{4.2}}\big)\tanh\big[\frac{3T_CM_{TD}(T)}{2TS(S+1)M_{4.2}}\big]$
= $M_{\rho}(T,M_{4.2})$ and $S(S+1) -
S^2\big(\frac{M_{K}(T)}{M_{4.2}}\big)^2 -
S\big(\frac{M_{K}(T)}{M_{4.2}}\big)$ = $M_{\rho}(T,M_{4.2})$
respectively. Consequently, one can actually compare and analyze
the $M_{\alpha}(T)/M_{4.2}$ ($\alpha$ = TD, K, $\rho$) calculated
from Tinbergen-Dekker (TD), Kasuya (K) and Eq.~(\ref{eq:7}) with
the experimentally measured $M_{exp}(T)/M_{4.2}$. However, one has
to switch to Eq.~(\ref{eq:9}) given below for the hole-doped
strongly correlated non-ferromagnetic semiconductors, which is
again based on iFDS~\cite{arulsamy6,andrew},

\begin{eqnarray}
\rho_h =
\frac{A_h(m^*_h)^{\frac{-1}{2}}}{2e^2}\left[\frac{2\pi\hbar^2}{k_B}\right]^{3/2}T^{1/2}
\exp\left[\frac{E_I + E_F}{k_BT}\right]. \label{eq:9}
\end{eqnarray}

\section{3. Discussion}\lb{s-in}

\subsection{3.1. Diluted magnetic semiconductors}\lb{s-in}

It is the purpose of this paper to explain the transport mechanism
below $T_C$ without violating the physical properties known in
$\rho(T>T_C)$. The discussion on the mechanism of transport
properties of ferromagnets is seen through the eyes of resistivity
measurements both in the presence of and in the absence of {\bf
H}. The resistivity measurements~\cite{esch1} and its fittings
based on Eqs.~(\ref{eq:7}) and~(\ref{eq:9}) are shown in
Fig.~\ref{fig:1} a) and b) respectively for Ga$_{1-x}$Mn$_x$As.
Literally, one needs two fitting parameters ($A$ and $E_I$) for
$\rho(T>T_C)$ and another two ($B$ and $M_{\rho}(T,M_{4.2})$) for
$\rho(T<T_C)$. All the fitting parameters are given in
Table~\ref{Table:I}. Note that $S$ = 1 and 5/2 are employed for
the fittings of $M_K(T)/M_{4.2}$ while $T_C$ and $T_{cr}$ were
determined from the experimental resistivity curves, {\it not}
from the magnetization measurements or any other techniques. The
deviation of $M_K(T)/M_{4.2}$ from the $M_{exp}/M_{4.2}$ increases
with $S$ from 1 $\to$ 5/2. The $\rho(T)$ is found to increase with
$x$ from 0.060 to 0.070 due to the mechanism proposed by Van Esch
{\it et al.}~\cite{esch1,esch11} and Ando {\it et
al.}~\cite{ando12}. They proposed that neutral Mn$^{3+}$ acceptors
that contribute to magnetic properties could be compensated by As,
where for a higher concentration of Mn, instead of replacing Ga it
will form a six-fold coordinated centers with As
(Mn$^{6As}$)~\cite{esch1,esch11,ando12}. These centers will
eventually reduce the magnitude of ferromagnetism (FM) in DMS due
to the loss of spin-spin interaction between Mn(3$d^{5})$ and $h$.
Parallel to this, Mn$^{6As}$ formation is substantial in such a
way that Mn$^{3+}$ ions do not substitute Ga$^{3+}$ ions.
Therefore, $\rho(T)$ will be influenced by Mn$^{6As}$ clusters,
defects and Ga-Mn-As phase simultaneously significantly in this
range of $x$. This is also indeed in fact in accordance with iFDS
based resistivity models since if one assumes Mn$^{2+}$ ($E_I$ =
1113 kJmol$^{-1}$) or Mn$^{3+}$ ($E_I$ = 1825 kJmol$^{-1}$)
substitutes Ga$^{3+}$ ($E_I$ = 1840 kJmol$^{-1}$), then $\rho(T)$
should further decrease~\cite{arulsamy6} with $x$, which is not
the case here. Thus, iFDS also suggests that Mn$^{2+}$ or
Mn$^{3+}$ do not substitute Ga$^{3+}$. Interestingly, the
$T_{cr}$s observed in Ga$_{0.940}$Mn$_{0.060}$As ($T_{cr}$ = 10 K,
annealed: 370$^o$C) and Ga$_{0.930}$Mn$_{0.070}$As ($T_{cr}$ = 12
K, as grown) are identical with the calculated values, where $E_I$
+ $E_F$ = 8 K and 12 K respectively. $E_I$ + $E_F$ is actually
equivalent to $T_{cr}$ because of its exponential contribution as
shown in Eq.~(\ref{eq:7}). The calculated carrier density using
$E_I$ + $E_F$ (8, 12 K), $m^*_h$ = rest mass and Eq.~(\ref{eq:6})
is 2.4 $\times$ 10$^{19}$ cm$^{-3}$. Below $T_C$, spin alignments
enhance the contribution from $J_{se}$ and reduces the exponential
increase of $\rho(T)$. This reduction in $\rho(T)$ is as a result
of dominating $J_{se}$ and small magnitude of $E_I$ + $E_F$ (8-12
K), consequently its effect only comes around at low $T$ as
clearly shown in Fig.~\ref{fig:1} a). The
Ga$_{0.930}$Mn$_{0.070}$As samples after annealing at 370 $^o$C
and 390 $^o$C do not indicate any FM~\cite{esch1}
(Fig.~\ref{fig:1} b)). Thus the fittings are carried out with
Eq.~(\ref{eq:9}) that only require two parameters namely, $A$ and
$E_I$ + $E_F$ since $J_{se}$ = 0 (there is no observable $T_C$)
and/or $dM_{\alpha}(T)/M_{4.2}dT$ = 0 ($M_{\rho}(T,M_{4.2})$ =
constant). The exponential increase of $\rho(T)$ is due to $E_I$ +
$E_F$ as given in Eq.~(\ref{eq:9}) with zilch $J_{se}$
contribution.

Figure~\ref{fig:1} c) and d) indicate the calculated normalized
magnetization, $M_{\alpha}(T)/M_{4.2}$ obtained from
Eq.~(\ref{eq:7}). Note that $M_{\rho,TD,K}(T)/M_{4.2}$ is a
fitting parameter that has been varied accordingly to fit
$\rho(T<T_C)$. As a matter of fact, $M_{\rho}(T,M_{4.2})$ is used
to calculate $M_{\rho,TD,K}(T)/M_{4.2}$ with $S$ = 1.
$M_{\rho,TD,K}(T)/M_{4.2}$ is also compared with the
experimentally determined~\cite{esch1} $M_{exp}(T)/M_{4.2}$ as
depicted in Fig.~\ref{fig:1} d). One can easily notice the
relation, $M_{TD}(T)/M_{4.2}$ $>$ $M_{K}(T)/M_{4.2}$ $>$
$M_{\rho}(T)/M_{4.2}$ $>$ $M_{exp}(T)/M_{4.2}$ from
Fig.~\ref{fig:1} c) and d). As such, $M_{\rho}(T)/M_{4.2}$
determined from Eq.~(\ref{eq:7}) is the best fit for the
experimentally measured $M_{exp}(T)/M_{4.2}$. Obviously, the
higher number of aligned spins as calculated from Eq.~(\ref{eq:7})
using $M_{\rho}(T)/M_{4.2}$ compared to $M_{exp}(T)/M_{4.2}$ is
due to the ability of both $J_e$ and $J_{se}$ to follow the
easiest path. Simply put, resistivity measures only the path with
relatively lowest $E_I$ and with easily aligned spins that
complies with the principle of least action. In contrast,
magnetization measurement quantifies the average of all the spins'
alignments. On the other hand, the discrepancies of
$M_{TD}(T)/M_{4.2}$ with $M_{exp}(T)/M_{4.2}$ and
$M_{K}(T)/M_{4.2}$ with $M_{exp}(T)/M_{4.2}$ are due to long range
FM hole-hole and AFM Mn-hole interactions~\cite{esch1} apart from
the ability of both $J_e$ and $J_{se}$ to follow the easiest path.
The violation between $M_{TD,K}(T)/M_{4.2}$ and
$M_{exp}(T)/M_{4.2}$ suggests that the spin disorder scattering
alone is inadequate, in which, the principle of least action have
had played an enormous role.

\subsection{3.2. Manganites}\lb{s-in}

Now switching to manganites, Mahendiran {\it et
al}.~\cite{mahendiran2} discussed $\rho(T<T_C)$ with respect to
Eq.~(\ref{eq:1}) and obtained the activation energy, $E_a$ = 0.16
eV for $x$ = 0.1 and 0.2 of La$_{1-x}$Ca$_x$MnO$_3$ samples at 0
T. Using Eq.~(\ref{eq:7}) however, $E_I$ + $E_F$ for the former
and latter samples are calculated to be 0.12 and 0.11 eV
respectively. The calculated carrier density using $E_I$ + $E_F$
(0.12, 0.11 eV), $m^*_h$ = rest mass and Eq.~(\ref{eq:6}) is
approximately 10$^{17}$ cm$^{-3}$. In the presence of {\bf H} = 6
T, $E_I$ + $E_F$ is computed as 0.0776 eV for $x$ = 0.2 that
subsequently leads to $p$ = 10$^{18}$ cm$^{-3}$. It is proposed
that the activated behavior for $\rho(T>T_C)$ is due to $E_I$,
Coulomb interaction between ion and electron or rather due to the
polaronic effect~\cite{arulsamy6,arulsamy9}. The fittings are
shown in Fig.~\ref{fig:2} a) and b) while its fitting parameters
are listed in Table~\ref{Table:I}.
Theoretically~\cite{arulsamy6,arulsamy9}, Ca$^{2+}$ ($E_I$ = 868
kJmol$^{-1}$) $<$ La$^{3+}$ ($E_I$ = 1152 kJmol$^{-1}$), therefore
$\rho(T)$ is expected to decrease with Ca$^{2+}$ doping
significantly. Contradicting to that, only a small difference of
$E_I$ + $E_F$ between $x$ = 0.1 (0.12 eV) and 0.2 (0.11 eV) is
observed due to Mn$^{4+}$'s compensation effect where the quantity
of Mn$^{4+}$ increased 6\% from $x$ = 0.1 (19\%) to 0.2
(25\%)~\cite{mahendiran2}. To clearly see this, the difference of
$E_I$ between Ca$^{2+}$ and La$^{3+}$ is calculated, which is 1152
$-$ 868 = 284 kJmol$^{-1}$ and subsequently it is compared with
the 6\% increment of Mn$^{3+ \to 4+}$ ($E_I$ = 4940 kJmol$^{-1}$),
which is 0.81(1825) + 0.19(4940) $-$ 0.75(1825) $-$ 0.25(4940) =
187 kJmol$^{-1}$. Consequently, the actual difference is only 284
$-$ 187 = 97 kJmol$^{-1}$ instead of 284 kJmol$^{-1}$. This simple
calculation exposes that Ca$^{2+}$'s contribution has been
compensated with 6\% additional Mn$^{4+}$. All the values of $E_I$
discussed above were averaged in accordance with $E_I [X^{z+}] =
\sum_{i=1}^z\frac{E_{Ii}}{z}$ and should not be taken literally
since those $E_I$s are not absolute values. The absolute values
need to be obtained from the $r_B$ dependent $E_I$ equation stated
earlier. Prior to averaging, the 1$^{st}$, 2$^{nd}$, 3$^{rd}$ and
4$^{th}$ ionization energies for all the elements mentioned above
were taken from Ref.~\cite{winter}.

As is well known, at 6 T, La$_{0.8}$Ca$_{0.2}$MnO$_3$ indicate a
much lower resistivity (Fig.~\ref{fig:2} b)). The result that
larger {\bf H} giving rise to conductivity at $T$ $>$ $T_C$ is due
to relatively large amount of aligned spins at higher $T$ or {\bf
H} gives rise to $J_{se}$ at a higher $T$. Hence, $T_C$ at 6 T $>$
$T_C$ at 0 T and one can also conclude, $r_B$ (at 6 T) $>$ $r_B$
(at 0 T) due to the inequality, $E_I$ + $E_F$ = 78 meV (at 6 T)
$<$ $E_I$ + $E_F$ = 112 meV (at 0 T) complying with
Eq.~(\ref{eq:7}) and iFDS. Figure~\ref{fig:2} c) and d) depict
calculated $M_{\alpha}(T)/M_{4.2}$ with $S$ = 1 and
$M_{exp}(T)/M_{4.2}$ for $x$ = 0.2. Calculated $M_{TD}(T)/M_{4.2}$
is dropped for La$_{1-x}$Ca$_x$MnO$_3$ since $M_K(T)/M_{4.2}$
seems to be a better approximation than $M_{TD}(T)/M_{4.2}$ as
indicated in Fig.~\ref{fig:1} c) and d). Millis {\it et
al}.~\cite{millis14} have shown theoretically that double exchange
mechanism (DEM) alone is inadequate and one needs to incorporate
polaronic effect into DEM. In fact, Eq.~(\ref{eq:7}) takes both
polaronic effect ($E_I$) and DEM ($M_{\alpha}(T)/M_{4.2}$) into
account and yet there is a discrepancy between
$M_{\rho}(T)/M_{4.2}$ and $M_{exp}(T)/M_{4.2}$ though
Eq.~(\ref{eq:7}) reproduces $\rho(T)$ at all $T$ range accurately.
Again, this incompatibility is due to the principle of least
action as stated earlier. Note that the discrepancy between
$M_K(T)/M_{4.2}$ and $M_{\rho}(T)/M_{4.2}$ indicate the inadequacy
pointed out by Millis {\it et al}.~\cite{millis14}, not the
former. This inadequacy may not be clear from Fig.~\ref{fig:2} d)
because the deviation is large with respect to magnetization in a
narrow range of $T$ (careful observation will reveal this though).
In addition, the manganites' charge transport mechanism below
$T_C$ is also in accordance with Eq.~(\ref{eq:7}) because the
term, $M_{\rho}(T,M_{4.2})$ handles the exchange interactions'
complexities separately for DMS and manganites. For example, one
can clearly notice the different type of discrepancies between DMS
and manganites by comparing the empirical function,
$M_{\alpha}(T)/M_{4.2}$ ($\alpha$ = $\rho$, exp) between
Fig.~\ref{fig:1} d) and Fig.~\ref{fig:2} d). Hence,
Eq.~(\ref{eq:7}) is suitable for both types of ferromagnets, be it
diluted or concentrated. In conclusion, the resistivity model that
have incorporated the polaronic and magnetization effects has been
derived based on iFDS and spin disorder scattering theories. This
model is able to explain the transport mechanism below the Curie
temperature without violating the physical properties above $T_C$.
The discrepancy of the magnetization curves calculated from this
resistivity model with the experimental data of DMS arises as a
result of the ability of both spin and charge currents to follow
the easiest path according to the principle of least action, long
range ferromagnetic and short range antiferromagnetic
interactions. Whereas for manganites, this discrepancy is still
noticeable solely because of the principle of least action.

\section*{Acknowledgments}

The author is grateful to Arulsamy Innasimuthu, Sebastiammal
Innasimuthu, Arokia Das Anthony and Cecily Arokiam of CMG-A for
their financial assistances. ADA also thanks Bryne J.-Y. Tan,
Jasper L. S. Loverio and Hendry Izaac Elim for their kind help
with the extraction of experimental data points as given in Figure
1a) and 1b) as well as for providing some of the references. ADA
thanks Prof. Feng Yuan Ping for his support to participate in
ICMAT-2003, Singapore.

\begin{table}

\caption {Calculated values of $T$ independent electron-electron
scattering rate constant ($A$), $B$, which is a function of $T$
independent spin disorder scattering rate constant and spin
exchange energy ($J_{ex}$) as well as the ionization energy
($E_I$). Note that $T_C$ and $T_{cr}$ were obtained from the
resistivity curves. All these parameters are for Mn doped
Ga$_{1-x}$Mn$_x$As (as grown and annealed at 370 $^o$C, 390 $^o$C)
and Ca doped La$_{1-x}$Ca$_x$MnO$_3$ (measured at 0 and 6 T)
systems. All Ga$_{1-x}$Mn$_x$As samples were measured at 0 T.}
\label{Table:I}
\end{table}

\begin{figure}

\caption {Equation~(\ref{eq:7}) has been employed to fit the
experimental $\rho(T)$ plots for Ga$_{1-x}$Mn$_x$As as given in a)
whereas Eq.~(\ref{eq:9}) is used to fit the plots in b). All
fittings are indicated with solid lines. b) is actually for
annealed non-ferromagnetic Ga$_{0.930}$Mn$_{0.070}$As samples. c)
and d) show the $T$ variation of calculated
$M_{\alpha}(T)/M_{4.2}$ ($\alpha$ = K, TD, $\rho$) with $S$ = 1
for $x$ = 0.060 and 0.070 respectively. $M_K(T)/M_{4.2}$ is also
calculated with $S$ = 5/2 to indicate the type of deviation one
should expect with increasing $S$. The experimental
$M_{exp}(T)/M_{4.2}$ plots is only available for $x$ = 0.070 (as
grown) as shown in d).} \label{fig:1}
\end{figure}

\begin{figure}

\caption {Experimental plots of $\rho(T)$ for
La$_{1-x}$Ca$_x$MnO$_3$ at $x$ = 0.1, 0.2 and 0.2 (6 T) have been
fitted with Eq.~(\ref{eq:6}) as depicted in a) and b). All
fittings are indicated with solid lines. Whereas c) and d) show
the $T$ variation of calculated $M_{\alpha}(T)/M_{4.2}$ ($\alpha$
= K, $\rho$) with $S$ = 1 for $x$ = 0.1 and 0.2 respectively. The
experimental $M_{exp}(T)/M_{4.2}$ plots is only available for $x$
= 0.2, which is given in d).} \label{fig:2}
\end{figure}

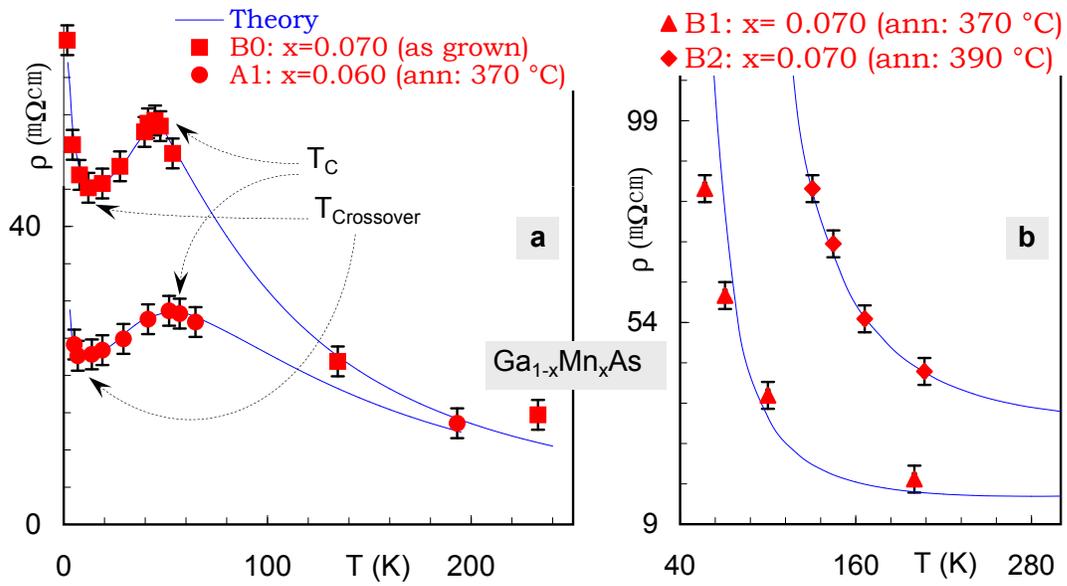
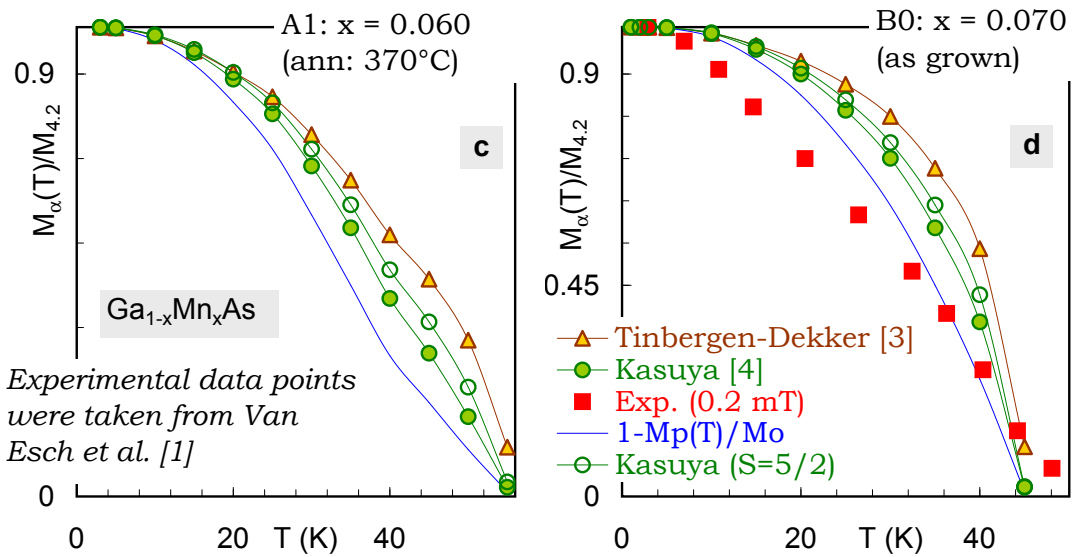
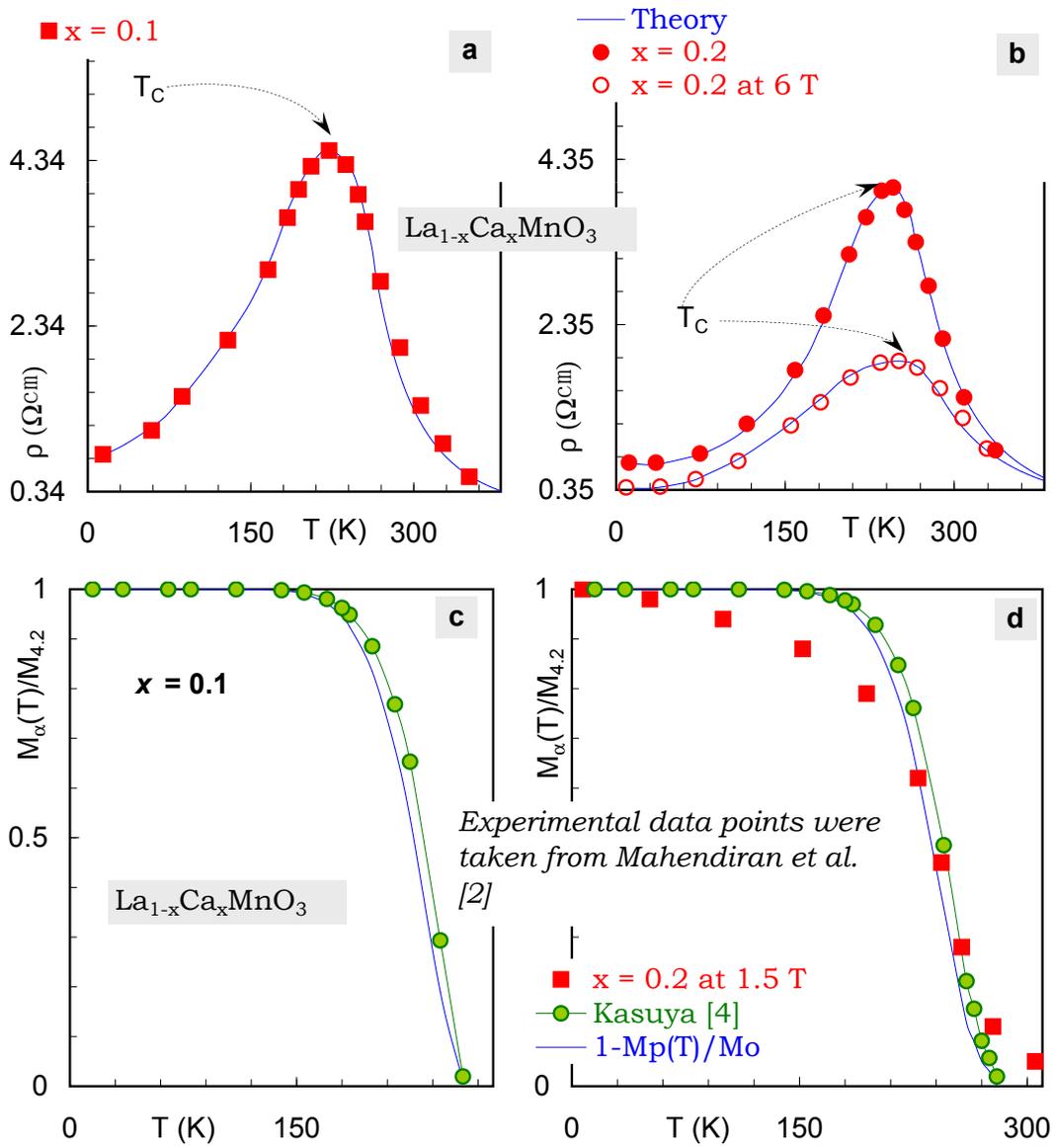
\end{document}